\documentclass[reprint,superscriptaddress,amsmath,amssymb,aps,prb,showkeys,showpacs]{revtex4-1}
\usepackage{graphicx}
\usepackage{tikz}
\usetikzlibrary{snakes}
\usepackage{xcolor}

\newcommand{\ket}[1]{\mathinner{|{#1}\rangle}}
\newcommand{\bra}[1]{\mathinner{\langle{#1}|}}

\newcommand{\boundellipse}[3]
{(#1) ellipse (#2 and #3)
}  

\begin{document}
\title{Vibrational Effects on the Formation of Quantum $W$ States}
\author{H. G. Mendon\c{c}a}
\affiliation{Instituto de F\'isica, Universidade Federal de Uberl\^andia, 38400-902 Uberl\^andia, MG, Brazil}
\author{F. M. Souza}
\email{fmsouza@ufu.br}
\affiliation{Instituto de F\'isica, Universidade Federal de Uberl\^andia, 38400-902 Uberl\^andia, MG, Brazil}
\date{\today}

\begin{abstract}
We theoretically investigate the formation of $W$ states in a tripartite system composed of three charge qubits coupled to vibrational modes. The electromechanical coupling is responsable for 
second order virtual processes that result in an effective electron-electron 
interaction between neighbor qubits, which yields to the formation of $W$ states.
Based on the Lang-Firsov transformation and perturbation theory,
we analytically solve the quantum dynamics, providing a mathematical 
expression for the maximally entangled $W$ state.
Dephasing is also taken into accout, paying particular attention 
on the robustness of bipartite entanglement against local dephasing processes.
\end{abstract}

\maketitle

\section{Introduction}
\label{sec:intro}

Entanglement is one of the main features of quantum mechanics that make quantum computers
so advantageous compared to classical computers.\cite{barnett2009} 
Entangled states appeared early in the quantum mechanics development 
in the context of the Einstein, Podolsky and Rosen (EPR) paradoxy.\cite{einstein1935}
Since then quantum entanglement became a prominent resource for quantum communication
and quantum information processing,\cite{chuang2004} with potential
applications in problems such as prime factoring\cite{lucero2012} and quantum simulations.\cite{qiu2020}

In the context of solid-state system, superconducting-based quantum devices 
have received great attention in the last decade, as they
constitute one of the leading system for implementation of quantum computation.\cite{devoret2013}
For instance, transmon qubits in superconducting chips have been used to generate 
the Greenberg-Horner-Zeilinger state.\cite{barends2014} The GHZ state\cite{greenberger2007}
was also found in a three-qubit superconducting circuit,\cite{dicarlo2010}
and both GHZ and W\cite{dur2000} states were reported in superconducting phase qubits.\cite{neeley2010}
Superconducting qubits have also been applied for quantum network.\cite{yin2015}
Additionally, a recent experiment with 53-qubits is an outstading example of the recent progress 
of the superconducting-based quantum computation.\cite{arute2019}

Even though superconducting qubits have received a great deal of attention and significant
progresses have been achieved, there are many other potential solid-state devices that
can be applied to manipulate qubits. For instance, it was demonstrated coherent oscillations
of single electron spin in GaAs quantum dots,\cite{koppens2005} and
coherent control of coupled electron spins was reported.\cite{petta2005} 
Additionally, it was demonstrated initialization, control and readout of three-electron spin qubits.\cite{medford2013}
More recently, silicon-based systems have being receved a growing attention
as it was found long electron spin coherence times.\cite{tyryshkin2012, maune2012, veldhorst2014, kawakami2014}
For instance, in the context of silicon based system it was recently reported
the implementation of CNOT gates and single-qubit operations.\cite{veldhorst2015}
It was also demonstrated an efficient resonantly driven CNOT gate for electron spins in silicon
double quantum dot structure.\cite{zajac2018}
In addition, SWAP two-qubit exchange gate between
phosphorus donor electron spin qubits in silicon was recently reported.\cite{he2019}
Also, semiconductor quantum dot system have been proposed as
solid state devices to construct GHZ state.\cite{nogueira2020} Recently, it was shown
that highly entangled two-qubit states can be achieved due to electron-vibrational mode coupling
in molecular systems.\cite{souza2019} Here we extend this previous work by showing that 
electromechanical coupling can also be a useful tool to generate entangled three-qubit 
$W$ states in electronic solid-state devices.

The $W$ states consist of a special class of entangled state, being of the form\cite{dur2000}
\begin{equation}
 \ket{W} = a \ket{001} + b \ket{010} + c \ket{100}, \phantom{x}
\end{equation}
($|a|^2+|b|^2+|c|^2 = 1$) which when one of the qubits is traced out leaves a partially entangled pair
of qubits. This is an example of a two-way entangled state,\cite{wong2001}
that is robust against losses in one of the qubits.\cite{dur2000}
Being of great importance for quantum information processing, such as 
two-party quantum teleportation,\cite{jung2008} and superdense coding,
the $W$ state have been both theoretically and experimentally investigated in an array of superconducting
microwave resonator,\cite{gangat2013} in spin systems,\cite{li2015, chen2017}
in superconducting quantum interference device.\cite{kang2016} Recently, 
three-photon $W$ states were demonstrated in optical fibers.\cite{fang2019}

In the present work we investigated a tripartite system composed of three charge qubits that interact 
with vibrational degrees of freedom of a nearby molecular structure,
that can be, for instance, carbon nanotubes as pointed out in Ref.
[\onlinecite{souza2019}]. Based on this previous work, here we derive a model Hamiltonian
based on the Lang-Firsov transformation, that explicitly shows a charge-charge attractive like 
interaction (a typical superconductivity related phenomena), 
that is behind the formation of the $W$ state being generated. Based on perturbation theory  a simple three dimensional model can be derived, expressed in the reduced computational base $\{\ket{001}, \ket{010}, \ket{100}\}$, that recovers the main physical ingredients behind our full numerical results. 
Our paper is organized as follows, in Sec. II we present our theoretical model and analytical
results, in Sec. III we show the quantum dynamics of our system, paying particular attention 
to the formation of the $\ket{W}$ state. In Sec. IV we account for dephasing processes and finally in Sec. V we summarize our conclusions.

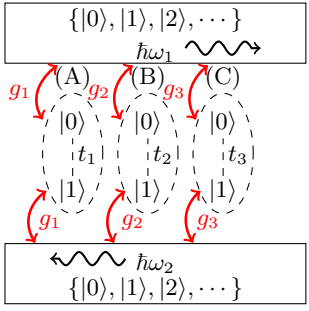
\begin{figure}[tb] 
\begin{center}
\begin{tikzpicture} [scale=1.0]

  \draw (0.4, 3.6) rectangle (4.4, 2.8);
  \draw (2.4,2.9) node[above=0.2cm] {$\{|0\rangle, |1\rangle,|2\rangle,\cdots \}$};
  \draw (2.4,2.5) node[above=0.2cm] {$\hbar \omega_1 $};
  \draw[->,snake=coil,segment aspect=0,black,thick]  (2.8,3.) -- (3.8,3.);
  
  \draw (0.4, -0.4) rectangle (4.4, 0.4);
  \draw (2.4,-0.3) node[above=0.2cm] {$\hbar \omega_2 $};
  \draw (2.4,-0.68) node[above=0.2cm] { $\{|0\rangle, |1\rangle,|2\rangle,\cdots \}$};
  \draw[->,snake=coil,segment aspect=0,black,thick]  (2.,0.2) -- (1.,0.2);
  
  \draw[]  (1.3,2.28) node[above=0.05cm] {(A)};
  \draw[]  (1.5,1.3) node[above=0.05cm] {$t_1$};
  \draw[dashed] (1.3,1.3) -- (1.3,1.8); 
  \draw[] (1.3,1.1)  node[] {$|1\rangle$}; 
  \draw[] (1.3,2.) node[] {$|0\rangle$};
  \draw[dashed] \boundellipse{1.3,1.6}{0.4}{0.8}; 
  
  \draw[]  (2.3,2.28) node[above=0.05cm] {(B)};
  \draw[]  (2.5,1.3) node[above=0.05cm] {$t_2$};
  \draw[dashed] (2.3,1.3) -- (2.3,1.8); 
  \draw[] (2.3,1.1)  node[] {$|1\rangle$}; 
  \draw[] (2.3,2.) node[] {$|0\rangle$};
  \draw[dashed] \boundellipse{2.3,1.6}{0.4}{0.8};

  \draw[]  (3.3,2.28) node[above=0.05cm] {(C)};
  \draw[]  (3.5,1.3) node[above=0.05cm] {$t_3$};
  \draw[dashed] (3.3,1.3) -- (3.3,1.8); 
  \draw[] (3.3,1.1)  node[] {$|1\rangle$}; 
  \draw[] (3.3,2.) node[] {$|0\rangle$};
  \draw[dashed] \boundellipse{3.3,1.6}{0.4}{0.8};
  
  \draw[line width=1,red, <->] (0.9,2.05) .. controls +(120:0.5cm) and +(50:0cm) .. (1.1,2.77); 
  \draw[line width=1,red, <->] (1.94,2.05) .. controls +(120:0.5cm) and +(50:0cm) .. (2.1,2.77); 
  \draw[line width=1,red, <->] (2.94,2.05) .. controls +(120:0.5cm) and +(50:0cm) .. (3.1,2.77); 

  \draw[red] (0.6,2.) node[above=0.2cm] {$g_1 $}; 
  \draw[red] (1.65,2.) node[above=0.2cm] {$g_2 $};
  \draw[red] (2.65,2.) node[above=0.2cm] {$g_3 $}; 
  
  \draw[line width=1,red, <->] (0.8,0.41) .. controls +(120:0.5cm) and +(50:0cm) .. (.95,1.1); 
  \draw[line width=1,red, <->] (1.85,0.41) .. controls +(120:0.5cm) and +(50:0cm) .. (2.,1.1); 
  \draw[line width=1,red, <->] (2.9,0.41) .. controls +(120:0.5cm) and +(50:0cm) .. (3.,1.1); 

  \draw[red] (1.,0.27) node[above=0.2cm] {$g_1 $}; 
  \draw[red] (2.1,0.27) node[above=0.2cm] {$g_2 $};
  \draw[red] (3.1,0.27) node[above=0.2cm] {$g_3 $}; 
  
\end{tikzpicture}
\end{center}
\caption{Illustration of the considered system. Three charge qubits
couple to two bosonic environments. No direct coupling between
the qubits are considered, so the qubits indirectly couple
to each other via vibrational modes of the bosonic subsystems.
As a result, the electrons in the qubits experience an attractive interaction, that gives
rise to the formation of highly entangled $\ket{W}$ state.}
\label{fig1}
\end{figure}

\section{Theoretical Model}
\label{sec:model}

Consider a multipartite system composed of five subspaces,
$\mathcal{H}=\mathcal{H}_{qb1}\otimes\mathcal{H}_{qb2}\otimes\mathcal{H}_{qb3}\otimes\mathcal{H}_{v}\otimes\mathcal{H}_{v}$, where the first three subspaces correspond to 
the qubits A, B and C, while the last two subspaces are associated to the vibrational modes, 
as illustrated in Fig. (\ref{fig1}). 
The qubits Hamiltonian is given by
\begin{equation}\label{Hdot}
 H_{\mathrm{qubits}} = \bigoplus_{n=1}^3 [ \frac{\delta_n}{2}\sigma_z^{(n)} + t_n \sigma_x^{(n)} ],
\end{equation}
where the Kronecker sum means $\bigoplus_{1}^{3} a_n A^{(n)}= a_1 A \otimes I \otimes I + a_2 I \otimes A \otimes I + a_3 I \otimes I \otimes A$. Also, $\delta_n$ and $t_n$ are the detuning between the electronic levels
and the intra qubit charge tunneling, respectively. The vibrational modes degree of freedom are described according to
\begin{equation}\label{Hv}
 H_{\mathrm{v}} = \omega_1 B^\dagger B \otimes I_v + \omega_2 I_v \otimes B^\dagger B,
\end{equation}
where $\omega_i$ is the energy of the $i$-th vibrational mode.
The operator $B$ ($B^\dagger$) annihilates (creates) vibrational excitations,
and $I_v$ is the identity matrix in the vibrational subspace.
The electromechanical coupling is given in terms of projection operators as 
\begin{eqnarray}\label{V}
 V &=& \bigoplus_{n=1}^3 [g_n P_0^{(n)}] \otimes [(B^\dagger+B) \otimes I_v] +\nonumber\\
    && \bigoplus_{n=1}^3 [g_n P_1^{(n)}] \otimes [I_v \otimes (B^\dagger+B)],
\end{eqnarray}
where $P_i=\ket{i}\bra{i}$ and $g_n$ is the coupling parameter between the electronic and vibrational degrees of freedom. 
This means that state $\ket{0}$ couples to the vibrational mode
characterized by $\omega_1$, 
while state $\ket{1}$ couples to the vibrational model with $\omega_2$. \cite{footnote1}
This feature is illustrated in Fig. (\ref{fig1}).
As recently shown in Ref. [\onlinecite{souza2019}], to deal with this particular model (which considers the electron-vibrational mode coupling) it is more convenient to use the Lang-Firsov canonical transformation.\cite{mahan2000} Defining the operator
\begin{eqnarray}
S &=& \bigoplus_{n=1}^3 [\lambda_n P_0^{(n)}] \otimes [(B^\dagger-B) \otimes I_v] +\nonumber\\
  && \bigoplus_{n=1}^3 [\lambda_n P_1^{(n)}] \otimes [I_v \otimes (B^\dagger-B)],
\end{eqnarray}
where $\lambda_n = g_n / \omega_n$, we can perform the transformation
$\bar{H}=e^{S} H e^{-S}$. After a straightforward calculation we find the following set of equations,
\begin{eqnarray} \label{Hqubits}
\bar{H}_{\mathrm{qubits}} &=& \bigoplus_{n=1}^3 \frac{\delta_n}{2}\sigma_z^{(n)}  -\frac{1}{\omega}(g_1^2+g_2^2+g_3^2) \nonumber\\
  &-&\frac{g_1 g_2}{\omega} (\sigma_z \otimes \sigma_z \otimes I + I^{\otimes 3} )\nonumber \\
  &-&\frac{g_1 g_3}{\omega} (\sigma_z \otimes I \otimes \sigma_z + I^{\otimes 3} )\nonumber\\
  &-&\frac{g_2 g_3}{\omega} (I \otimes \sigma_z \otimes \sigma_z + I^{\otimes 3} ).\nonumber\\
\end{eqnarray}
Notice that the last three terms of Eq.(\ref{Hqubits}) can be associated to an attractive Coulomb like interaction. 
The vibrational modes remains
as $\bar{H}_{\mathrm{v}} = \omega_1 B^\dagger B \otimes I_v + \omega_2 I_v \otimes B^\dagger B$.
Finally, the interaction term takes the form
\begin{eqnarray}
\bar{V} &=&  t_1 (\sigma_+ \otimes I \otimes I \otimes D(\lambda_1) \otimes D(-\lambda_1) + h.c.)+\nonumber\\
  && t_2 (I \otimes \sigma_+ \otimes I \otimes D(\lambda_2) \otimes D(-\lambda_2) + h.c.)+\nonumber\\
  && t_3 (I \otimes I \otimes \sigma_+ \otimes D(\lambda_3) \otimes D(-\lambda_3) + h.c.),
\end{eqnarray}
where $\sigma_+ = |0 \rangle \langle 1|$, $\sigma_- = |1 \rangle \langle 0|$ and $D(\pm \lambda_i) = e^{\pm \lambda_i (B^\dagger - B)}$ is the displacement operator.\cite{scully1997}  
In what follows we assume $t_i=t$, $\omega_i=\omega$ and $g_i=g$, so that $\lambda_i=\lambda=g/\omega$.

\begin{figure}[tb]
\centering\includegraphics[width=0.95\linewidth]{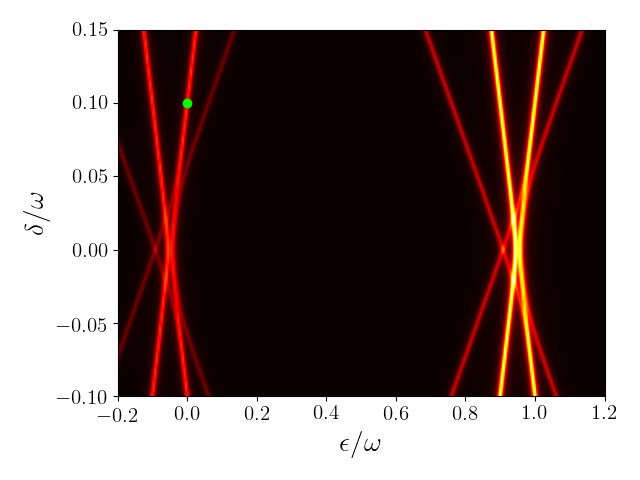}
\caption{Spectral function of the system as function of energy $\epsilon$ and detunning $\delta$. For small
values of energy we have a set of four branches, 
corresponding to Eqs. (\ref{e000})-(\ref{e110}) with $m=l=0$.
The two low intensity branches are given by $\varepsilon_{000,00} = (3/2) \delta$ and 
$\varepsilon_{111,00} = - (3/2) \delta$, while the other two high intensity branches are due to
the degenerate energies
$\varepsilon_{001,00} = \varepsilon_{010,00} = \varepsilon_{100,00} = (1/2) \delta$ and 
$\varepsilon_{011,00} = \varepsilon_{101,00} = \varepsilon_{110,00} = - (1/2) \delta$.
For high values of energy we find replicas of these branches as the energy of the vibrational modes
increases with $(n+l)\omega$. 
To form a $\ket{W}$ state within subspace $\mathcal{E}_\mathrm{w}$ spanned by $\{\ket{001}, \ket{010}, \ket{100}\}$
we need to set the energy parameters in such a way that 
the corresponding energies $\varepsilon_{001,00}$, $\varepsilon_{010,00}$ and $\varepsilon_{100,00}$ 
are well isolated from the other energy levels. The green dot in the figure indicates the 
set of energies configuration used in our simulation to obtain a $\ket{W}$ 
state within the aforementioned subspace. Virtual processes take place between 
the many branches shown in order to construct this state.}
\label{fig2}
\end{figure}

From the transformed model, we can easily find the system eigenenergies 
in the absence of tunneling, i.e., $\bar{V}=0$.
In what follows, this coupling term will be considered as a perturbation.
The eigenenergies of $\bar{H}_0 = \bar{H}_{\mathrm{}}+\bar{H}_{\mathrm{v}}$ can be divided
in four energy groups. The higher and lower energetic ones,
\begin{equation}
 \varepsilon_{000,ml} = \frac{1}{2} (\delta_1 + \delta_2 + \delta_3) - \frac{3g^2}{\omega} +(m+l)\omega, \label{e000}
\end{equation}
and
\begin{equation}
\varepsilon_{111,ml} = -\frac{1}{2} (\delta_1 + \delta_2 + \delta_3) - \frac{3g^2}{\omega} +(m+l)\omega, \label{e111}
\end{equation}
respectively. The upper intermediate ones,
\begin{eqnarray}
\varepsilon_{001,ml} &=& \frac{1}{2} (+\delta_1 + \delta_2 - \delta_3) + \frac{g^2}{\omega} +(m+l)\omega,\\ \label{e001}
\varepsilon_{010,ml} &=& \frac{1}{2} (+\delta_1 - \delta_2 + \delta_3) + \frac{g^2}{\omega} +(m+l)\omega,\\ \label{e010}
\varepsilon_{100,ml} &=& \frac{1}{2} (-\delta_1 + \delta_2 + \delta_3) + \frac{g^2}{\omega} +(m+l)\omega, \label{e100}
\end{eqnarray}
and the lower intermediate ones,
\begin{eqnarray}
\varepsilon_{011,ml} &=& \frac{1}{2} (+\delta_1 - \delta_2 - \delta_3) + \frac{g^2}{\omega} +(m+l)\omega,\\ \label{e011}
\varepsilon_{101,ml} &=& \frac{1}{2} (-\delta_1 + \delta_2 - \delta_3) + \frac{g^2}{\omega} +(m+l)\omega,\\ \label{e101}
\varepsilon_{110,ml} &=& \frac{1}{2} (-\delta_1 - \delta_2 + \delta_3) + \frac{g^2}{\omega} +(m+l)\omega. \label{e110}
\end{eqnarray}
A constant energy shift of $-6g^2/\omega$ was omitted in all the energies above.
If we set the detunings at $\delta_1=\delta_2=\delta_3=\delta$ we find
$\varepsilon_{000,ml} =  \frac{3}{2} \delta - \frac{3g^2}{\omega} +(m+l)\omega$,
$\varepsilon_{111,ml} = -\frac{3}{2} \delta - \frac{3g^2}{\omega} +(m+l)\omega$,
and the degenerate levels 
$\varepsilon_{001,ml}=\varepsilon_{010,ml}=\varepsilon_{100,ml}=\frac{1}{2} \delta + \frac{g^2}{\omega} +(m+l)\omega$, 
and $\varepsilon_{011,ml}=\varepsilon_{101,ml}=\varepsilon_{110,ml}= -\frac{1}{2} \delta + \frac{g^2}{\omega} +(m+l)\omega$. 
In order to have a graphical view of these energy levels we calculate the spectral function
of the system. Consider the retarded Green function,\cite{jauho2008}
\begin{equation}
 G_{n n'}^r (t-t') = -i \theta(t-t') \bra{\phi_n} e^{-i \bar{H}_0 (t-t')} \ket{\phi_{n'}}.
\end{equation}
Assuming $\ket{\phi_n}$ as the eigenstate of $\bar{H}_0$, $\bar{H}_0 \ket{\phi_n} = \varepsilon_{n} \ket{\phi_n}$,
we can write
\begin{equation}
 G_{n n'}^r (t-t') = -i \theta(t-t') \delta_{n n'} e^{-i \varepsilon_{n} (t-t')}.
\end{equation}
Fourier transforming $G_{n n'}^r (t-t')$ we find
\begin{equation}
 G_{n n'}^r (\epsilon) = \frac{\delta_{n n'}}{\epsilon - \varepsilon_{n} + i \eta},
\end{equation}
where $\eta \to 0^+$. The spectral function, defined as 
$A_n (\epsilon) = -2 \mathrm{Im} [G_{nn}^r (\epsilon)]$, is then given by
\begin{equation}
 A_n (\epsilon) = \frac{2 \eta}{(\epsilon-\varepsilon_n)^2 + \eta^2}.
\end{equation}
In Fig. (\ref{fig2}) we show the total spectral function $A(\epsilon)=\sum_n A_n (\epsilon)$ as function of 
$\epsilon$
and $\delta$ for $t_n=0$. We observe two spaced groups of four branches each.
The high energetic ones are basically replicas of the lowest ones,
corresponding to vibrational modes $n=1$ and $m=0$ or $n=0$ and $m=1$.
Focusing on the low energetic levels ($n=0$, $m=0$) we have 
$\varepsilon_{000,ml} =  \frac{3}{2} \delta - \frac{3g^2}{\omega}$ and
$\varepsilon_{111,ml} = -\frac{3}{2} \delta - \frac{3g^2}{\omega}$,
with high slope $3/2$ (in modulus) as $\delta$ increases.
The other two branches are given by 
$\varepsilon_{001,00}=\varepsilon_{010,00}=\varepsilon_{100,00}=\frac{1}{2} \delta + \frac{g^2}{\omega}$
and $\varepsilon_{011,00}=\varepsilon_{101,00}=\varepsilon_{110,00}= -\frac{1}{2} \delta + \frac{g^2}{\omega}$.
If $\delta$ is properly tunned such that the levels 
$\varepsilon_{001,00}$, $\varepsilon_{010,00}$ and $\varepsilon_{100,00}$
are relatively far from the other branches, the system becomes suitable for the
generation of $\ket{W}$ states within subspace $\mathcal{E}_\mathrm{w}= \mathrm{span}\{ \ket{100}, \ket{010}, \ket{001} \}$.
It is important to emphasize that to form $W$ states within
this electronic subpsace, the physical parameters should be set along the branches 
corresponding to the energies $\varepsilon_{001,00}=\varepsilon_{010,00}=\varepsilon_{100,00}$. 
Without loss of generality, we set our parameters to the values as indicated by the green dot in 
Fig. (\ref{fig2}). Other points could also be chosen, as long as they satisfy
the condition of being far apart from other branches. With this assumption, it is convenient to
divide the system into two subspaces, the relevant one, given by the
projector operator $P = (\ket{100}\bra{100} + \ket{010}\bra{010} + \ket{001}\bra{001})\otimes\ket{00}\bra{00}$,
and the irrelevant one $Q=I-P$, composed by all the other states in the computational basis.
In order to estimate the effective coupling between the states in $P$ subspace,
we apply the second order perturbation theory. For instance, we can calculate
the coupling between states $\ket{100}$ and $\ket{010}$, 
\begin{eqnarray}
 \Omega_{100,010} &=& - \sum_{i,j,k} \frac{\bra{100}\bar{V}\ket{ijk}\bra{ijk}\bar{V}\ket{010}}{\varepsilon_{ijk,00}-\varepsilon_{100,00}}, 
\end{eqnarray}
where we consider only $n=m=0$ for the vibrational states, as high energetic levels contributions to the sum are neglected. More specifically we have,
\begin{eqnarray}
 \Omega_{100,010} &=& - \frac{\bra{100}\bar{V}\ket{000}\bra{000}\bar{V}\ket{010}}{\varepsilon_{000,00}-\varepsilon_{100,00}} \nonumber \\
                  && - \frac{\bra{100}\bar{V}\ket{110}\bra{110}\bar{V}\ket{010}}{\varepsilon_{110,00}-\varepsilon_{100,00}},
\end{eqnarray}
which results in,
\begin{equation}
  \Omega_{100,010} = - \frac{4 g^2 t^2 e^{-2\lambda^2}}{\omega \delta (\delta - \frac{4 g^2}{\omega})},
\end{equation}
where the identity $\bra{m} D(\lambda) \ket{0} = \lambda^m e^{-\lambda^2/2}/\sqrt{m!}$ was applied.\cite{scully1997} 
Similar results hold for $\Omega_{100,001}$ and $\Omega_{010,001}$. Therefore, we can write
an effective model in the subspace $P$ as $H_{eff}=\Omega h$, where
\begin{equation}
 h = \ket{100}\bra{010}+\ket{100}\bra{001}+\ket{010}\bra{001}+H.c.
\end{equation}
The eigenvalues of $h$ are given by $a'=2$, $b'=c'=-1$, with corresponding eigenvectors
\begin{eqnarray}
 \ket{\bar{0}'} &=& \frac{1}{\sqrt{3}} (\ket{\bar{0}}+\ket{\bar{1}}+\ket{\bar{2}}) \\
 \ket{\bar{1}'} &=& \frac{1}{\sqrt{6}} (\ket{\bar{0}}+\ket{\bar{1}}-2\ket{\bar{2}}) \\
 \ket{\bar{2}'} &=& \frac{1}{\sqrt{2}} (\ket{\bar{0}}-\ket{\bar{1}}).
\end{eqnarray}
From here on, we take the shorthand notation $\ket{\bar{0}}=\ket{100}$, $\ket{\bar{1}}=\ket{010}$ and $\ket{\bar{2}}=\ket{001}$. With this simple model we can proceed to the dynamics analysis.

\emph{Physical Parameters.} In the experimental point of view, 
our model can describe carbon nanotube (CNT) quantum dots, as recently
proposed in Ref. [\onlinecite{souza2019}] for two qubits entanglement.
In the context of CNT the vibrational
mode frequency will be assumed around $\omega = 20$ meV, in agreement with typical values
found in CNT.\cite{leroy2004} The intra qubit electron hopping will be considered $t = 0.1$ meV, in accordance
with the coupling parameter found in experimental setup on parallel CNT quantum dots.\cite{gob2013}
Also, the electron-phonon coupling parameter $g$ can be experimentally adjusted in CNT,\cite{benyamini2014}
here in particular we set $g = 0.1\omega$. Different $g$ values were explored in the context of quantum transport 
in CNT quantum dots.\cite{walter2013, sowa2017} Even though we assume 
experimentally feasible parameters in the context CNT quantum dots, the theoretical results presented
below can in principle be found in different system, as soon as the system set of parameters
matches the conditions discussed in Fig. (\ref{fig2}).

\section{Quantum Dynamics}

In this section we calculate the evolution of the quantum state, searching for 
the formation of quantum entanglement between the three qubits. Consider
\begin{equation}
 \ket{\psi(\phi)} = e^{-i h \phi} \ket{\psi_0},
\end{equation}
where $\phi = \Omega t$, with $\ket{\psi_0}=\ket{\bar{0}}$ as initial state.
Writing the relevant states $\ket{\bar{0}}$, $\ket{\bar{1}}$ 
and $\ket{\bar{2}}$ in terms of the $h$ eigenstates, we have
\begin{eqnarray}
 \ket{\bar{0}} &=& \frac{1}{\sqrt{3}} \ket{\bar{0}'} + \frac{1}{\sqrt{6}} \ket{\bar{1}'} + \frac{1}{\sqrt{2}} \ket{\bar{2}'} \\
 \ket{\bar{1}} &=& \frac{1}{\sqrt{3}} \ket{\bar{0}'} + \frac{1}{\sqrt{6}} \ket{\bar{1}'} - \frac{1}{\sqrt{2}} \ket{\bar{2}'}  \\
 \ket{\bar{2}} &=& \frac{1}{\sqrt{3}} \ket{\bar{0}'} - \frac{2}{\sqrt{6}} \ket{\bar{1}'}.
\end{eqnarray}
Applying the evolution operator on $\ket{\psi_0}$ we find
\begin{equation}
 \ket{\psi(\phi)} = \frac{e^{-i 2 \phi}}{\sqrt{3}} \ket{\bar{0}'} + \frac{e^{i \phi}}{\sqrt{6}} \ket{\bar{1}'} + 
 \frac{e^{i \phi}}{\sqrt{2}} \ket{\bar{2}'},
\end{equation}
which in the original computational basis reduces to
\begin{equation}\label{psitheoretical}
 \ket{\psi(\alpha)} = [ \cos(\alpha) + \frac{i}{3} \sin(\alpha) ] \ket{\bar{0}} - \frac{2 i}{3} \sin(\alpha) (\ket{\bar{1}}+\ket{\bar{2}}),
\end{equation}
with $\alpha=(3/2)\phi$.
The maximum entangled state can be found at 
\begin{equation}\label{alphamax}
 \sin(\alpha_{max}) = \pm \sqrt{3}/2
\end{equation}
so that $\alpha_{max} = \pm \frac{\pi}{3} + \pi n$, or
\begin{equation}
 \frac{\alpha_{max}}{2\pi} = \pm \frac{1}{6} + \frac{n}{2}.
\end{equation}
This provides the following sequence of integer numbers
\begin{equation}\label{sequence}
 \beta_{\mathrm{max}} \equiv \frac{6 \alpha_{max}}{2\pi} = 1, 2, 4, 5, 7, 8, 10, 11, 13, ...,
\end{equation}
that gives the times with maximum entangled states.
At this point it becomes instructive to compare our analytical predictions with
a numerical calculation derived from the full Hamiltonian given by Eqs.(\ref{Hdot})-(\ref{V}). 
Following W. D\"ur \emph{et al.},\cite{dur2000} we calculate the quantity
\begin{equation}
 E_{\tau} \equiv \mathcal{C}_{AB}^2 + \mathcal{C}_{AC}^2 + \mathcal{C}_{BC}^2,
\end{equation}
where $\mathcal{C}_{AB}$ is the concurrence\cite{wootters1998} for the reduced density operator $\rho_{AB}=\mathrm{Tr}_{C}\{\rho\}$. Similar definition holds for $\mathcal{C}_{AC}$ and $\mathcal{C}_{BC}$.
It was shown that $E_{\tau}$ attains its larger value $4/3$ for $W$ state.\cite{dur2000} 
We also calculate the quantity,
\begin{equation}
 \mathcal{C}_{\min}^2 \equiv \min{(\mathcal{C}_{AB}^2, \mathcal{C}_{AC}^2, \mathcal{C}_{BC}^2)},
\end{equation}
which was proved to be $\mathcal{C}_{\min}^2 = 4/9$ for the $W$ state.\cite{dur2000}
In Fig. \ref{fig3}(a)-(b) we show both $E_{\tau}$ and $\mathcal{C}_{\min}^2$
as a functio of $\beta$. We clearly see that these two quantities peak
at $\beta$ values given by the integer numerical sequence in Eq. (\ref{sequence}).
Additionally, the peaks become close to the corresponding expected values,
$4/3$ and $4/9$ (horizontal red lines).
This indicates that our analytical predictions agree with the full numerical calculations. The present result is
a proof of concept that electromechanical resonator can be a
useful tool to generate highly entangled $W$ states.

We can also analytically find the exact quantum $W$ state generated at $\beta_{\mathrm{max}}$ values.
Using Eq. (\ref{alphamax}) in Eq. (\ref{psitheoretical}) we obtain
\begin{equation}\label{psialphamaxl}
 \ket{\psi(\alpha_{\mathrm{max}})} = \frac{1}{\sqrt{3}}
 [\ket{\bar{0}} + e^{\mp i \frac{2\pi}{3}}(\ket{\bar{1}}+\ket{\bar{2}})],
\end{equation}
where global phases are omitted. Defining the target density matrix
\begin{equation}
 \sigma_{\mathrm{tar}} = \ket{\psi(\alpha_{\mathrm{max}})} \bra{\psi(\alpha_{\mathrm{max}})} \otimes \ket{00}\bra{00},
\end{equation}
we can calculate the fidelity
\begin{equation}\label{fidelity}
 \mathcal{F} = \mathrm{Tr}\{ \sigma_{\mathrm{tar}} \rho(t) \}.
\end{equation}
In Fig. \ref{fig3}(c) we show $\mathcal{F}$ as a function of $\beta$, for both signs 
of the relative phase $e^{\mp i \frac{2\pi}{3}}$. Notice that $\mathcal{F}$ becomes
close to one at the $\beta$ values given by Eq. (\ref{sequence}), alternating 
the relative phase sign.

Finally, in Fig. \ref{fig3}(d) we show the concurrences $C_{BC}$ (black line) and $C_{AB}$ (red line) against $\beta$.
The peaks of these concurrences indicate the formation of bipartite entanglement.
In particular, the peaks of $C_{BC}$ appear above $0.8$, thus indicating a relative high
entanglement degree. Notice that we can minimize the probability amplitude of state $\ket{\bar{0}}$ in 
Eq. (\ref{psitheoretical}), in order to find the corresponding state for the bipartite subsystem $BC$.
This amplitute cannot reach zero, though its minimum value can result in partially entanglement
between qubits $B$ and $C$. Writing this probability as
\begin{equation}
 p(\alpha) = \cos^2(\alpha) + \frac{1}{9} \sin^2(\alpha),
\end{equation}
and taking the derivative with respect to $\alpha$ we
find the condiction $\sin(\alpha) \cos(\alpha) = 0$.
The minimum is reached for $\cos(\alpha) = 0$, thus $\alpha=\pi/2, 3\pi/2, 5\pi/2, ...$.
Expressing these values in terms of $\beta$ we find the following sequence
\begin{equation}\label{betasequence}
 2\beta= 3, 9, 15, ...,
\end{equation}
which perfectly agrees with the peaks of $C_{BC}$ (black solid lines) in Fig. 3(d).
This shows that the system tends to partially entangle qubits $B$ and $C$.

\section{Dephasing}
\label{sec:dephasing}

In this last section we discuss how dephasing mechanisms taken in
one of the three qubits can affect the entanglement of the other two qubits.
To do so, we assume that phase flip erros are present
in qubit A. As pointed out in Ref. [\onlinecite{dur2000}] the entanglement of
the $\ket{W}$ state is more robust against loss mechanisms in
one of the qubits, when compared to other states, e.g., the $\ket{GHZ}$ state.
Now we see this effect in action in our modeled device. The Lindblad equation is written in its
standard form as\cite{lindblad1976}
 \begin{eqnarray}\label{rhodephasing}
 \dot{\rho} &=& -i[H,\rho] + \sum_{i} (L_i \rho L_i^\dagger - \frac{1}{2} L_i^\dagger L_i \rho - \frac{1}{2} \rho L_i^\dagger L_i),
\end{eqnarray}
where the Lindblad operator is given by $L_i = \sqrt{\Gamma} \sigma_z \otimes I \otimes I$ for $i=A$,
with $\Gamma$ being a phase flip error rate. Here we take $\Gamma = 1 \cdot 10^{-4} \omega$,
which corresponds to 0.5 GHz for $\omega = 20$ meV.
We also assume $L_B = L_C = 0$, i.e., Bob and Claire's qubits are assumed free of
dephasing.

\begin{figure}[t]
    \centering
    \includegraphics[width=0.5\textwidth]{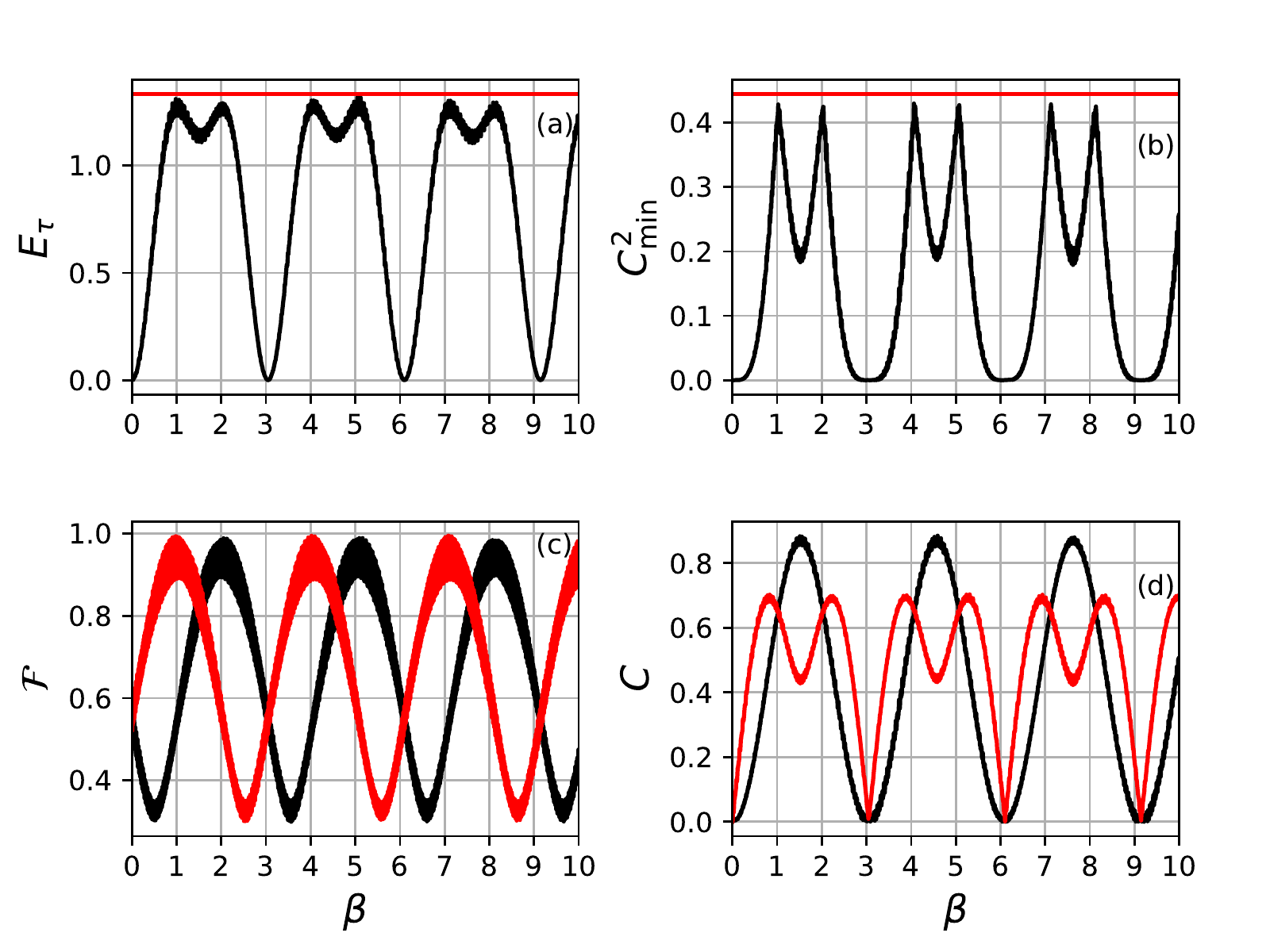}
    \caption{Dynamics of the quantities $E_{\tau}$, $\mathcal{C}_{\min}^2$, fidelity and concurrence as a function of $\beta$. In panels (a) and (b) we show $E_{\tau}$ and $C_{min}^2$. Notice that both curves peak around $\beta=1,2,4,5,7,8...$, getting close to their maximum values $4/3$  and $4/9$ (red line), respectively. This suggests the formation of maximum entangled $|W\rangle$ state at these particular $\beta$ values.
    In panel (c) we show the fidelity for the target state $|\psi(\alpha_{\mathrm{max}})\rangle$ with both phase signs 
    $e^{ i \frac{2\pi}{3}}$ (red curve) and $e^{- i \frac{2\pi}{3}}$ (black curve).    
    Note that $\mathcal{F} \approx 1$ at the same $\beta$ values where $E_\tau$ and $C_{min} ^2$ peaks occur, thus revealing the $|W\rangle$ states being formed.     In panel (d) we show both the concurrences between qubits $BC$ ($C_{BC}$, black line) and qubits $AB$ ($C_{AB}$, red line). 
    The peaks observed for $C_{AB}$ and $C_{BC}$ indicate the formation of partially bipartite entanglement. For instance, the peaks of $C_{BC}$ (black line) suggest the formation of a partially entangled state between qubits $B$ and $C$. Analogously, the peaks of $C_{AB}$ (red line) indicate the formation of partial entanglement between qubits $A$ and $B$. Parameters: $g = 0.1 \omega$, $\delta = 0.1 \omega$ and $t = 0.005 \omega$.}
    \label{fig3}
\end{figure}

\begin{figure}[t]
    \centering
    \includegraphics[width=0.5\textwidth]{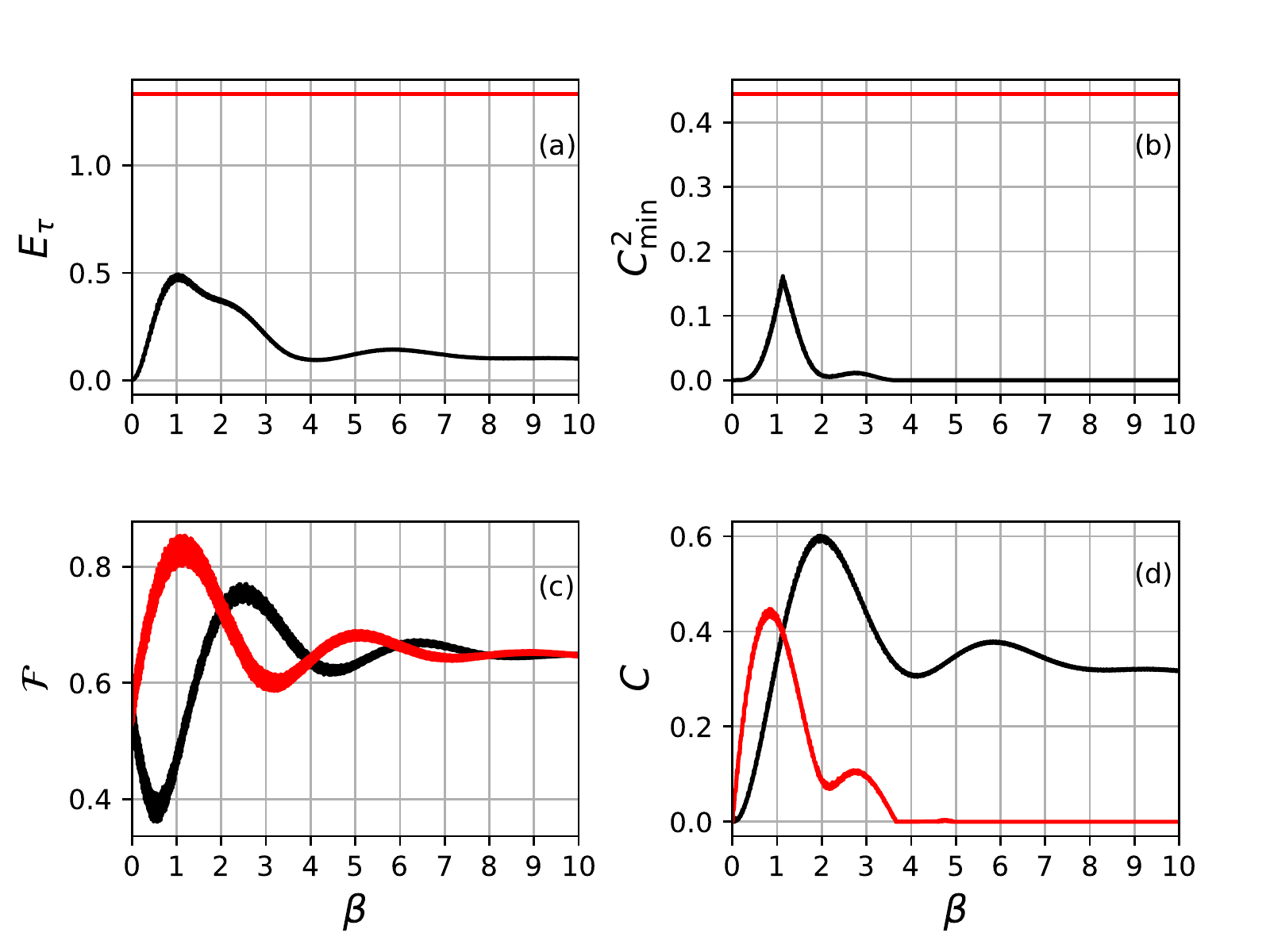}
    \caption{Dynamics of $E_{\tau}$, $\mathcal{C}_{\min}^2$, fidelity and concurrences against $\beta$ in the presence of dephasing in qubit $A$.
    The quantities (a) $E_{\tau}$ and (b) $C^2 _{min}$ are suppressed due to dephasing, indicating that tripartite entanglement is quite sensible to dephasing in one of the three qubits.
    The fidelity in panel (c) shows damped oscillation with only the first peak becoming close to the target $|W\rangle$ state.
    In panel (d) we show both the concurrences $C_{AB}$ (red line) and $C_{BC}$ (black line). 
    As a result of dephasing in qubit $A$, the concurrence $C_{AB}$ is suppressed.
    In contrast, $C_{BC}$ is preserved for longer times. 
    Parameters: $g = 0.1 \omega$, $\delta = 0.1 \omega$ and $t = 0.005 \omega$, $\Gamma = 1 \cdot 10^{-4} \omega$ ($0.5$GHz).}
    \label{fig4}
\end{figure}

Similarly to Fig. (\ref{fig3}), in Fig. (\ref{fig4}) we show the quantities
$E_{\tau}$ and $\mathcal{C}_{\min}^2$, the fidelity $\mathcal{F}$ and concurrences
for the bipartite subsystems $C_{AB}$ (red line) and $C_{BC}$ (black line).
The peaks of both $E_{\tau}$ and $\mathcal{C}_{\min}^2$ are significantly suppressed,
thus departing from the respective ideal values of $4/3$ and $4/9$. The fidelity now 
shows damped oscillations. Interestingly, though, the concurrences $C_{AB}$ and $C_{BC}$
present contrasting features, with $C_{BC}$ remaining finite much longer
than $C_{AB}$. This shows that the subsystem composed of 
qubits $B$ and $C$ preserves for relatively large times some degree of entanglement even when dephasing is present in qubit $A$.

\section{Conclusion}
\label{sec:tcl}

We investigate how the coupling between charge degrees of freedom
and vibrational modes can yield to the formation of
highly entangled $\ket{W}$ states. 
We numerically calculate quantities
such as $E_{\tau}$ and $\mathcal{C}_{\min}^2$, that peak on the formatio of $\ket{W}$ states. 
Based on perturbation theory and the Lang-Firsov transformation, 
we derive analytical expressions that recover our full numerical calculation, 
thus providing a simple physical picture of the complex dynamics driven by the
many body interaction. In particular, we analytically predict with
great accuracy the times in which the $\ket{W}$ state is being formed.
Additionally, via concurrence calculation we find partial entanglement between 
subsystems composed of only two electronic qubits, i.e., tracing out one of the qubits. 
We show that even in the presence of dephasing mechanisms taken in one of the qubits, a partial
entanglement between the other two qubits is preserved for relatively large times, thus revealing the
robustness of the present system against local dephasing processes.

\begin{acknowledgments}
This work was supported by Conselho Nacional de Desenvolvimento 
Cient\'ifico e Tecnol\'ogico (CNPq).
\end{acknowledgments}

\appendix

\section*{Supplemental Material}

The model presented in Eqs.(\ref{Hdot})-(\ref{V}) is suitable in the description of
quantum computation and quantum information, such as calculation of entanglement properties, 
as the subspaces and qubits structures are emphasized.
However, if we are interested in quantum transport phenomena, it becomes more suitable to
deal with second quantization formalism. The second quantization allows the inclusion
in the computational basis states such as $\ket{vac}$, i.e., the vacuum state, 
that can be experimentally achieved by draining all the 
particles of the system into nearby attached electrodes. Alternatively, we can have a state
with all the electronic states being occupied due to charge injection from electrodes with high 
enough chemical potentials. In order to achieve such description 
we can rewrite our model Hamiltonian in its second quantized form
\begin{eqnarray}
 H_{dot} &=& \sum_{i=1}^6 \varepsilon_i d_i^\dagger d_i + (t_1 d_1^\dagger d_2 + t_2 d_3^\dagger d_4 + t_3 d_5^\dagger d_6 + h.c.), \nonumber\\
\end{eqnarray}
where $d_i$ ($d_i^\dagger$) annihilates (creates) one electron in quantum dot $i$. In this description
each qubit is composed by two quantum dots, i.e., qubit A: dots 1-2, qubit B: dots 3-4 and qubit C: dots 5-6.
Each quantum dot has a single energy level $\varepsilon_i$ and the dots in each qubit are coupled
to each other via a hopping parameter $t_i$. For the bosonic subsystem we have
\begin{equation}
 H_b = \omega_1 b_1^\dagger b_1 + \omega_2 b_2^\dagger b_2,
\end{equation}
with $\omega_i$ being the energy of the $i$-th vibrational mode and $b_i$ ($b_i^\dagger$) the annihilation
(creation) operator for the bosonic field. The electron-phonon coupling
can be written in its standard form,\cite{footnote2}
\begin{eqnarray}\label{V_second}
 V &=& g_1 d_1^\dagger d_1 (b_1^\dagger + b_1) + g_2 d_2^\dagger d_2 (b_2^\dagger + b_2) \nonumber \\
  &+& g_1 d_3^\dagger d_3 (b_1^\dagger + b_1) + g_2 d_4^\dagger d_4 (b_2^\dagger + b_2) \nonumber \\
  &+& g_1 d_5^\dagger d_5 (b_1^\dagger + b_1) + g_2 d_6^\dagger d_6 (b_2^\dagger + b_2) 
\end{eqnarray}
where $g_i$ gives the electron-phonon coupling strength. This equation is the second quantizad version
of Eq. (\ref{V}). Notice, though, that the operator in Eq. (\ref{V_second}) belongs to a much
larger Hilber space when compared to the one in Eq. (\ref{V}), 
as each quantum dot can sustain two states, namely, no particle $\ket{1}$ or one particle $\ket{0}$.
So for the electronic subspace we have $2^6$ possible states, contrasting to Eq. (\ref{V})
where the qubits description provides $2^3$ available electronic configurations.
The operators $d_i$ can be expressed in terms of Jordan-Wigner expansion, as described
in Ref. [\onlinecite{souza2017}]. 

Here we simulate an experimental scenario where the molecular system is initially empty of particles, 
i.e., its initial quantum state $\ket{n_1 n_2 n_3 n_4 n_5 n_6} \otimes \ket{N_1 N_2}$ is given by $\ket{111111}\otimes\ket{00}$,
with $n_i=1$ indicating no particle in $i$-th quantum dot and $N_i=0$ no excitation in the $i$-th 
bosonic bath. The quantum state that will evolve is then initialized with a pulse
of gate voltage that controls the charge injection in each quantum dot. This kind of
initialization can be experimentally implemented.\cite{shinkai2009}
To this aim we apply the following Lindblad equation,\cite{lindblad1976}
\begin{equation}\label{lindblad2}
 \dot{\rho} = -i [H_0, \rho] - \frac{1}{2} \sum_{i=1}^{6} \Gamma_i [ d_i d_i^\dagger \rho - d_i^\dagger \rho d_i + h.c.], 
\end{equation}
with $H_0 = H_{dot} + V + H_b$. Notice that this Lindblad equation differs from the one in Eq. (\ref{rhodephasing}). 
Here it accounts for charge injection from reservoirs (leads) into the quantum dots,
while in Eq. (\ref{rhodephasing}) it describes pure dephasing.
We assume that $\Gamma_i$ is governed by external gate
voltages that control the charge tunneling between leads and dots.
In particular, we assume that only $\Gamma_2$, $\Gamma_3$ and $\Gamma_5$ assume nonzero values,
with charges being injected in quantum dots $2$, $3$ and $5$. This gate pulse results in the following initialized
state
\begin{equation}
 \ket{\psi_0} = \ket{100101}\otimes\ket{00}.
\end{equation}
After the initialization pulse, the system evolves reaching highly entangled states.
Here in particular we will focus on the formation of the same states found in Eq. (\ref{psialphamaxl}),
that correspond to the peaks observed in Fig. \ref{fig3}(c). In the present notation
the target state can be written as
\begin{equation}\label{target2}
 \ket{\phi_{\pm}} = [\ket{100101} + e^{\pm i 2 \pi /3} (\ket{011001}+\ket{010110})]\otimes\ket{00}/\sqrt{3}.
\end{equation}
Defining the target state as $\sigma_{\mathrm{tar}} = \ket{\phi_{\pm}}\bra{\phi_{\pm}}$ we compute the fidelity
$\mathcal{F} = \mathrm{Tr}\{ \sigma_{\mathrm{tar}} \rho(t) \}$, with the density matrix provided by Eq. (\ref{lindblad2}).

\begin{figure}[t]
    \centering
    \includegraphics[width=0.5\textwidth]{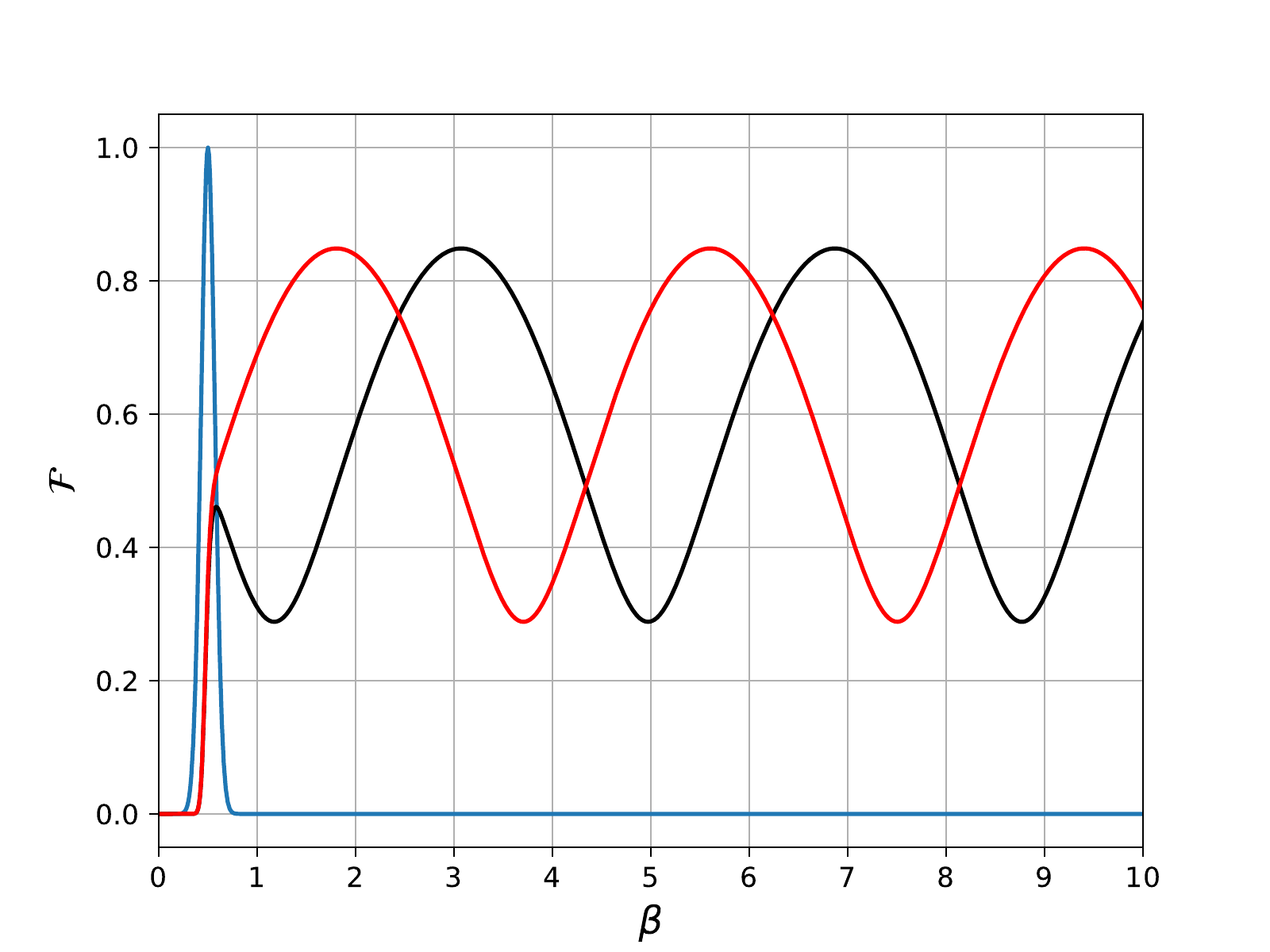}
    \caption{Fidelity against $\beta$ in the presence of a gate voltage pulse (blue line) that initializes the quantum state.
    Both signs of the relative phases of the target state $\ket{\phi}_{\pm}$ are considered, $e^{- i 2 \pi /3}$ (black line) and $e^{+ i 2 \pi /3}$
    (red line). The fidelity peak is reduced when compared to the result in Fig \ref{fig3}(c), due to the decoherence imposed
    by the charge injection occuring in the initialization gate pulse. Parameters: $g = 0.1 \omega$, $\delta = 0.1 \omega$ and $t = 0.005 \omega$.}
    \label{fig5}
\end{figure}

Figure (\ref{fig5}) shows $\mathcal{F}$ against $\beta$ after the system being initialized by
a gate pulse (blue line). Both phase signs are shown, $e^{- i 2 \pi /3}$ (black line) and $e^{+ i 2 \pi /3}$
(red line). Notice that these results are very similar to the ones found in Fig. \ref{fig3}(c).
The main contrast is the maximum value reached by the fidelity in the presence of initialization pulse,
which is lower when compared to the case illustrated in Fig. \ref{fig3}(c). This fidelity suppression
is related to the initialization procedure that imposes some decoherence to the system.
To inject charge into the quantum system we need to open it during some time interval, by connecting the molecular structure to 
charge reservoirs. This coupling between system and charge reservoirs (leads) brings unavoidable decoherences
that are reflected in a reduction of the fidelity degree in the subsequent quantum evolution.

\end{document}